\documentclass{nature}
\def\KVS{KV$_3$Sb$_5$}
\def\CVS{CsV$_3$Sb$_5$}

\def\AVS{AV$_3$Sb$_5$}
\def\cm{cm$^{-1}$}
\def\Tc{$T_{\rm DW}$}
\def\DCDW{$\Delta_{\rm DW}$}
\usepackage{graphicx}
\usepackage{amsmath}
\usepackage{multicol}
\usepackage{xcolor}
\usepackage{hyperref}
\hypersetup{
    colorlinks,
citecolor=blue,
linkcolor=red,
urlcolor=blue,
    }
\makeatletter
\let\saved@includegraphics\includegraphics
\AtBeginDocument{\let\includegraphics\saved@includegraphics}
\renewenvironment*{figure}{\@float{figure}}{\end@float}
\makeatother
\newcommand{\onlinecite}[1]{\hspace{-1 ex} \nocite{#1}\citenum{#1}}

\title{Optical detection of the density-wave instability in the kagome metal KV$_3$Sb$_5$}

\author{Ece Uykur$^{1,*}$, Brenden R. Ortiz$^{2,3}$, Stephen D. Wilson$^{2}$, Martin Dressel$^{1}$, Alexander A. Tsirlin$^{4,*}$}

\begin{document}

\maketitle

\begin{affiliations}
 \item 1.~Physikalisches Institut, Universit\"at Stuttgart, D-70569 Stuttgart, Germany
 \item Materials Department, University of California Santa Barbara, Santa Barbara, CA, 93106, United States
 \item California Nanosystems Institute, University of California Santa Barbara, Santa Barbara, CA, 93106, United States
 \item Experimental Physics VI, Center for Electronic Correlations and Magnetism, Augsburg University, 86159 Augsburg, Germany\\
email: ece.uykur@pi1.physik.uni-stuttgart.de; altsirlin@gmail.com
\end{affiliations}

\begin{abstract}
Coexisting density-wave and superconducting states along with the large anomalous Hall effect in the absence of local magnetism remain intriguing and enigmatic features of the \AVS\ kagome metals (A = K, Rb, Cs). Here, we demonstrate via optical spectroscopy and density-functional calculations that low-energy dynamics of \KVS\ is characterized by unconventional localized carriers, which are strongly renormalized across the density-wave transition and indicative of electronic correlations. Strong phonon anomalies are prominent not only below the density-wave transition, but also at high temperatures, suggesting an intricate interplay of phonons with the underlying electronic structure. We further propose the star-of-David and tri-hexagon (inverse star-of-David) configurations for the density-wave order in KV$_3$Sb$_5$. These configurations are strongly reminiscent of $p$-wave states expected in the Hubbard model on the kagome lattice at the filling level of the van Hove singularity. The proximity to this regime should have intriguing and far-reaching implications for the physics of KV$_3$Sb$_5$ and related materials.
\end{abstract}


\section*{Introduction}
Kagome geometry of corner-sharing triangles plays a special role in condensed-matter physics. In magnetic insulators, it may cause the highly entangled quantum spin liquid state with exotic fractionalized excitations\cite{Broholm2020}. Kagome metals exploit another aspect of this peculiar geometry, the simultaneous presence of linear and flat bands. The former cross at the Dirac point that gives rise to Weyl nodes when time-reversal symmetry is broken. These interesting features of the electronic structure were indeed detected in several magnetic kagome metals\cite{Lin2018,Ye2018,Liu2019,Yin2019,Kang2020}, where spin texture -- the type of magnetic order and its individual spin directions -- was shown to have large impact on the energy bands and transport\cite{Ye2019,Guguchia2020,Biswas2020}.

The discovery of \AVS{} (A = K, Rb, Cs) \cite{Ortiz2019} opens the way to studying physics of kagome framework in the absence of local magnetic moments. Here, the topological nature of the kagome bands can be intricately intertwined with electronic instabilities, such as superconductivity and charge order. Under ambient conditions, the \AVS\ family crystallizes in the hexagonal $P6/mmm$ space group. The V-atoms create the kagome network, while the Sb1 atoms fill the centers of the kagome triangles. These V-Sb1 layers are sandwiched between graphite-like Sb2 layers that are, in turn, separated by the alkali-metal ions (Fig.~\ref{1}a), creating easily exfoliable quasi-2D structures.  

The \AVS{} compounds show abrupt changes in the magnetic susceptibility and electrical resistivity around 78\,K (A = K\cite{Ortiz2019}), 102\,K (A = Rb\cite{Yin2021}), and 94\,K (A = Cs\cite{Ortiz2020a}). They further become superconducting below 0.9\,K (A = K\cite{Ortiz2021} and Rb\cite{Yin2021}) and 2.5\,K (A = Cs\cite{Ortiz2019}). On one hand, robust nature of the former anomaly under magnetic field and absence of local magnetism witnessed by muon spin spectroscopy\cite{Kenney2021} suggest that mostly charge degrees of freedom should be involved. On the other hand, the anomalous Hall effect is strongly enhanced in \KVS{}\cite{Yang2020} and \CVS{}\cite{Yu2021} below the 78\,K (94\,K) transition and has been ascribed to scattering from spin clusters. Furthermore, the recent STM results on \KVS\ demonstrate chiral response under reversed magnetic field\cite{Jiang2021} and also suggest that the low-temperature electronic state may not be a conventional charge order. This raises the question what this electronic state is, and in which form spin degrees of freedom can be embedded in it.

Theoretical studies of the kagome Hubbard model predict the
formation of unusual spin states in this setting. When the Fermi level is tuned to the van Hove singularity that arises from the band saddle point at $M$ (Fig.~\ref{1}b), different forms of density-wave order are expected\cite{Kiesel2013}. Their $p$-wave symmetry\cite{Nayak2000} implies that no spatial modulation of spin density takes place, and any magnetic order ensues from spin currents. Doping such states away from the van Hove singularity gives rise to an unconventional superconductivity with different gap symmetries\cite{Kiesel2013,Kiesel2012,Wang2013}. This reveals several interesting parallels to \AVS\ compounds where both the electronic transition and superconductivity have been reported.

Another aspect, which is hitherto absent in models of kagome metals but inevitably present in real materials, are phonons that may couple to the electronic background. Phonons are instrumental in driving charge-density-wave instabilities via strong momentum dependence of the electron-phonon coupling\cite{Johannes2008,Zhu2015}. Although density-wave instabilities of a model kagome metal may be purely electronic in nature and do not require phonons \textit{per se}, their manifestations in real \AVS\ materials seem to involve atomic displacements\cite{Ortiz2020a}. This renders phonons and electron-phonon coupling potentially important ingredients of the \AVS\ physics.

With these questions in mind, here we use broadband optical spectroscopy and density-functional calculations to examine the electronic structure of and the 78\,K anomaly in one of the \AVS\ members, namely, \KVS{}. We demonstrate that the anomaly manifests itself by a density-wave-like behavior in our optical probe. Using strong sensitivity of the infrared spectroscopy to charge degree of freedom, we identify bulk nature of the density-wave order and gauge the associated energy scale. The crossover to this density-wave state is accompanied by a strong renormalization of the unconventional carrier response that appears as a low-energy localization peak and reflects the importance of electronic correlations in this system. Furthermore, phonon anomalies observed above and below the transition indicate a strong coupling of phonons to the underlying electronic structure. Aided by \textit{ab initio} band-structure calculations, we propose possible structural models of the density-wave (DW) state and identify their similarity to phases predicted for the nearest-neighbor Hubbard model on the kagome lattice\cite{Kiesel2013}. The respective filling factor puts the Fermi level at the van Hove singularity, in strong resemblance to \KVS{} with its high density of states at the Fermi energy. We argue that similarities between the proposed DW structures and the predicted phases of the kagome Hubbard model can be the key to understanding the peculiar physics of \KVS.

\section*{Results}
\subsection{Normal state.}

Above 0.3~eV, optical conductivity of \KVS{} shows a rather frequency-independent behavior with several interband transitions, which get sharper upon cooling (see Supplementary Figure 2 for details). This frequency-independent behavior is interrupted with a very pronounced absorption in the low-energy range below 0.3~eV, as shown in Fig.~\ref{2}a.

A closer look reveals the details. The low-energy part of the spectra can be decomposed into several contributions as demonstrated in Fig.~\ref{2}b. A very sharp Drude component is present at low energies extrapolating to the dc resistivity values. Narrow peaks that do not shift in frequency upon cooling represent two phonon modes. A broader peak around 0.1~eV does not move either and can be associated with the low-energy interband transition, discussed in detail in the following paragraph. Finally, an even broader peak is observed around 0.06~eV at room temperature and displays a red shift upon cooling. We interpret this feature as a localization peak and defer its discussion to a later part. 

Band-structure calculations show that in \KVS\ and its sibling compounds the crossings of the linear Dirac bands lie below the Fermi level (see for example Fig.~\ref{2}e). This should lead to a spectral weight transfer from the interband to the intraband transitions below the Pauli-blocked edge and result in the sharp Drude component\cite{Gusynin2006}, which is indeed observed in our spectra. The small scattering rate of around 60-70\cm\ extracted from the fits of the optical conductivity is also in line with the high mobility of the carriers residing in these topologically non-trivial bands\cite{Biswas2020}.

The interband transitions are readily identified with the help of DFT calculations for the undistorted $P6/mmm$ crystal structure of \KVS{} (Fig.~\ref{1}e). Linear bands appear in the vicinity of the Fermi level and arise from the underlying kagome geometry of the V atoms (Fig.~1). Band-resolved calculation (Fig.~\ref{2}c) reveals that in the $0.25-1.0$\,eV range the optical conductivity is dominated by two nearly frequency-independent contributions that are indeed expected for transitions between linear bands in 2D Dirac systems\cite{Gusynin2006, Kuzmenko2008}. We assign these contributions to the transitions between the linear parts of bands A-C and B-D, respectively. 

The sharp absorption peak around 0.1~eV has a different origin and can be traced back to the transitions between bands B and C containing parallel segments along the $A-L$ reciprocal direction. We note in passing that the size of this peak is quite sensitive to the exact position of the Fermi level and may be also influenced by adding electronic correlations within DFT+$U$. Best agreement with the experimental spectrum was achieved upon shifting the Fermi level by $+64$\,meV in DFT, or upon adding $U=2$\,eV. These changes control the occupation of bands B and C around the $M$-point, see Supplementary note 6 for details.

Notwithstanding the complex nature of the real \KVS{} band structure, several similarities to the idealized bands of a nearest-neighbor kagome metal can be identified (Fig.~\ref{1}b). First, crossing points of the linear bands are located below the Fermi level, suggesting the effective filling factor above $\frac13$. Second, the Fermi level sits at the energy where bands A--C cross in the vicinity of the $M$-point. Their inflection points are remarkably similar to the band shape expected in the idealized kagome metal near its van Hove singularity at the filling factor of 5/12. Indeed, in its normal state \KVS{} also shows a high density of states at the Fermi level (Fig.~\ref{1}c). This suggests that the 78\,K instability in \KVS{} may be associated with an instability of a correlated kagome metal at its van Hove singularity\cite{Kiesel2013}.

\subsection{Density-wave state.}

Having identified general trends in the optical conductivity, we inspect the low-temperature behavior in more detail. A transport and magnetic anomaly is observed at \Tc$=78$\,K (Fig.~\ref{1}d). Its insensitivity to the magnetic field\cite{Ortiz2019} suggests an electronic rather than magnetic nature of this instability. Indeed, in the optical spectra we observed clear signatures of this transition. Although no band gap opens below \Tc, the decrease in the density of states at the Fermi energy is witnessed by the reduction in the low-energy optical conductivity, which is transferred toward higher energies as dictated by the optical sum rules. These observations prove the formation of a bulk DW state in \KVS{}\cite{Basov2011}. The phonon modes persist below \Tc, albeit with significant anomalies.

Difference spectra illustrate the transfer of spectral weight (Fig.~\ref{2}d). The low-energy part of $\sigma_1(\omega)$ suppressed below $\sim$0.1\,eV is recovered at higher energies with a maximum around $\sim$0.17\,eV. The energy scale of the spectral weight redistribution is gauged by \DCDW\ that can be taken as an order parameter. We employed the zero-crossing of the difference spectra to determine 2\DCDW. Its temperature dependence is given in Fig.~\ref{2}e. Alternatively, one can choose the maximum of the SW transfer as an analogy to ARPES measurements, resulting in different absolute values of \DCDW{} but essentially the same temperature dependence when scaled to the 10\,K value. This temperature dependence deviates from the mean-field behavior, $\Delta_{\rm mf} \approx \Delta(0)\sqrt{1-\frac{T}{T_c}}$ (close to \Tc), where we assume \Tc$=78$\,K and $\Delta(0) = 59$~meV. The obtained energy scale is of the same order of $\sim$0.03\,eV as determined recently by STM \cite{Jiang2021}. Note however that here we probe the bulk response, while STM is sensitive to the surface states only. 

To identify possible DW states in \KVS{}, we rely on the experimental constraint of the $2\times 2$ in-plane superstructure detected by scanning tunneling microscopy\cite{Jiang2021,Zhao2021} and single-crystal x-ray diffraction on \CVS{}\cite{Ortiz2020a}. Different low-symmetry configurations constructed within the $2\times 2$ supercell were allowed to relax, and two stable solutions were found (Fig.~\ref{3}). One of them is star of David-type charge order reminiscent of 1T-TaS$_2$\cite{Wilson1975,Fazekas1979,Wang2020a} and denoted as \textit{star} in the following. The other one involves two types of vanadium clusters, triangles and hexagons, and is labeled simply as \textit{hexagon} for brevity. In both cases, the V--V distances inside the clusters are below 2.71\,\r A, to be compared with 2.74\,\r A in the undistorted structure, whereas V--V distances between the clusters exceed 2.80\,\r A.

The hexagon and star structures lie lower in energy than the parent, undistorted structure by 2.3\,meV/f.u. and 1.1\,meV/f.u., respectively. These energy differences are only marginally affected by the choice of the exchange-correlation functional or adding the spin-orbit coupling, and all three structures are non-magnetic in DFT.  Further on, both hexagon and star types of charge order are compatible with transport properties of \KVS{}. The metallic behavior observed experimentally even below \Tc\ suggests that a sizable Fermi surface persists below the transition, and the DOS at the Fermi level does not vanish. Indeed, calculated band structures are metallic.  The changes in DOS (Fig.~\ref{3}d) between the undistorted and DW structures span the energy range of about 0.1\,eV in good agreement with \DCDW{} determined from optics (see also Fig.~\ref{1}c). 

One peculiar aspect of this density-wave state is its persistent metallic nature. No band gap opens below \Tc, although the DOS at the Fermi level is reduced. This illustrates the multi-band nature of \KVS{} (Fig.~\ref{1}e). The vanadium kagome bands with the saddle points around $M$ are prone to the formation of the DW state, whereas other bands crossing the Fermi level in the vicinity of $K$ and $\Gamma$ are dominated by Sb states and thus less affected by the transition. 

Optical conductivity allows a further verification of the DW structures. Experimentally, we observe that the peak due to interband absorption shifts from 0.1\,eV above \Tc\ to about 0.15\,eV below \Tc. Such a blue shift is well reproduced by DFT calculations for the DW states, as demonstrated in Fig.~\ref{3}e. Therefore, both hexagon and star structures are likely candidates for the DW state of \KVS{}. The main changes in the electronic structure below \Tc\ are captured by the atomic displacements in the DW states.

An important observation at this juncture is that both hexagon and star structures are also close analogues of the DW states predicted theoretically for the Hubbard model at the filling level of the van Hove singularity\cite{Kiesel2013}. These states are shown in Fig.~\ref{3} and classified as charge bond order and spin bond order or, respectively, as singlet and triplet $p$-wave states according to the symmetry of their order parameter\cite{Nayak2000}. The star structure is immediately recognized as charge bond order, whereas the hexagon structure may be tentatively associated with spin bond order, albeit with one caveat. Spin bond order is not a conventional spin-density wave where spatial modulation of spin density creates local magnetic moments on individual atoms. Instead, it entails a spatial modulation of spin current that can not be captured by DFT. The hexagon structure itself, as obtained by the DFT atomic relaxation, is of charge bond order type, too, and can be understood as an inverse star-of-David, because the short and long V--V bonds are merely inverted compared to the star structure. However, the resemblance of this structure to the spin bond order is far from accidental. The hexagon structure does not support spin-density waves and appears to be immune to conventional magnetism, whereas the star structure readily turns into a spin-density wave when magnetism is introduced on the DFT+$U$ level (see Supplementary Note 6 for details). 

The analogy to the $p$-wave states of the kagome Hubbard model is reinforced by the similar filling factors. The $p$-wave instabilities appear in the model near its van Hove singularity\cite{Kiesel2013}, whereas real band structure of \KVS{} lies close to this regime and shows a peak in the density of states around the Fermi level. On the other hand, atomic reconstruction in the DW state should involve phonons, in contrast to the Hubbard model where only electronic instabilities are at play. This dichotomy -- competing roles of Fermi surface nesting and phonons -- has sparked vivid debates for many of the density-wave states reported earlier.

In the 1D case, electronic instabilities and phonons come hand in hand\cite{Monceau2012}. Extending this picture to higher dimensions is usually based on the assumption that Fermi surface nesting occurs at quasi-1D portion of the Fermi surface, and the resulting electronic reconstruction is sufficient to drive a metal-insulator transition or cause drastic changes in carrier scattering. However, other effects may be significant too. For instance, strong electron-phonon coupling was discussed as the source of the charge density wave observed in metallic 2D dichalcogenides, 2$H$-NbSe$_2$\cite{Johannes2008, Calandra2009}, where strong momentum-dependent electron-phonon coupling can cause it even in the absence of nesting. The situation becomes even more complex in high-temperature cuprate superconductors, where neither weak Fermi surface nesting near the antinodal region\cite{Comin2014} nor the strong electron-phonon coupling\cite{Zhu2015} were found sufficient to explain the density-wave instability. Strong electronic correlations were then proposed as another crucial ingredient\cite{Fujita2014}.

The aforementioned effects may also contribute to the DW formation in \KVS{}. Their relative importance depends on the balance between electronic correlations and electron-phonon coupling. In the following, we discuss possible experimental fingerprints of these two major ingredients using optical data for \KVS{}.

\subsection{Phonon Anomalies.}

Experimental optical spectra reveal two phonon modes at 188~\cm\ and 482~\cm, where the latter one is a pronounced antiresonance. Calculated frequencies of $\Gamma$-point phonons allow the assignment of the lower mode to the IR-active in-plane E$_{1u}$ phonon (black line in Fig.~\ref{4}a). In contrast, the upper mode could not be reproduced, because the highest optical phonon is at 299~\cm. On the other hand, DFT calculations reveal two IR-active modes at 237~\cm\ (E$_{1u}$) and 241~\cm\ (A$_{2u}$) , fairly close to each other (red line in Fig.~\ref{4}a). This makes the high-energy mode observable either as a combination of these modes or as an overtone of one, possibly the 241~\cm\ mode, as the resonance frequency is exactly doubled. The overtone/combination modes typically appear via anharmonic effects that may simultaneously reduce the intensity of the fundamental modes and even make them invisible\cite{Du2013}. Below, we also show that the fundamental mode at 188~\cm\ is unusually broad by virtue of its coupling to the electronic background. A similar broadening of the fundamental modes around 240~\cm\ would make them invisible in our spectrum. Alternatively, the 482~\cm\ mode could be of plasmonic origin\cite{Grigorenko2012} or a non-$\Gamma$ phonon that appears in the IR spectrum due to phonon-plasmon interactions\cite{Lapointe2017}. The latter interpretation seems less likely, though, because recent phonon calculations\cite{Tan2021} put the upper boundary of single-phonon excitations around 300~\cm. On the other hand, harmonic approximation assumed in this calculation remains to be verified experimentally, especially in the light of the abnormally high atomic displacement parameter of K atoms\cite{Ortiz2019} that potentially indicates strong anharmonic effects in \KVS.

To analyze the aforementioned modes, we fit temperature-dependent spectra, as described in the Supplementary note 3. Both modes show striking signatures of the electron-phonon coupling. For the lower 188~\cm\ mode, strong anomalies are observed across the transition. The resonance frequency ($\omega_0$), intensity ($\Delta\epsilon$), and damping (linewidth, $\gamma$) obtained from the Lorentzian fit are given in Fig.~\ref{4}d-f. Below \Tc, a strong softening is accompanied by the significant increase in the intensity. Furthermore, the line is very broad and by far exceeds the resolution of our measurement (1~\cm).

Below \Tc, this broadening may be explained by the presence of multiple modes that are expected in the DW states based on our phonon calculations (Fig.~\ref{4}a). On the other hand, the undistorted structure shows only one mode between 100 and 200~\cm. Therefore, the broadening above \Tc\ must be caused by the coupling to the electronic background. Moreover, the line gets broader upon cooling, while the sharpening is usually observed when thermal fluctuations abate. Such an unexpected broadening of the phonon modes upon cooling has been discussed in terms of an electron-phonon coupling scenario\cite{Bonini2007, Lazzeri2006}, for instance in the case of graphene. Here we used the same formalism and fitted temperature dependence of the linewidth (Fig.~\ref{4}f) using 
\begin{equation}
\gamma^{\rm e-ph}(T) = \gamma^{\rm e-ph}(0)\left[f\left(-\frac{\hbar\omega_0}{2k_BT}\right)-f\left(\frac{\hbar\omega_0}{2k_BT}\right)\right]
\end{equation} 
Here, $\hbar\omega_0 = 174$\,\cm\ is the calculated $E_{1u}$ phonon energy, $k_B$ is the Boltzmann constant, and $f(x) = 1/[e^x+1]$. The intrinsic linewidth of $\gamma^{e-ph}(0) = 52$\,\cm\ signals the crucial role of electron-phonon coupling and makes it a plausible reason for the DW instability in \KVS\cite{Zhu2015}.

The high-energy mode displays a strong Fano-like behavior, indicating that it couples to the electronic background, too. The fitting parameters for this mode are given in Fig.~\ref{4}g-i. While the resonance frequency is nearly temperature-independent, the intensity ($\Delta\epsilon$) increases towards \Tc\ and decreases below the transition. On the other hand, the broadening ($\gamma$) increases continuously upon cooling. The coupling (asymmetry) constant $q$ suggests the stronger Fano-character at higher temperatures. On approaching \Tc, the mode becomes more symmetric, while below \Tc\ it can be represented by a completely symmetric antiresonance. Below, we argue that the peculiar evolution of this mode is strongly intertwined with changes in the localization peak that we discuss in the following.

\subsection{Localization peak and electronic correlations.}

Another distinct feature of \KVS{} is the presence of a localization peak in the low-energy part of the spectrum. This peak, sometimes understood as a displaced Drude peak, is identified by its red shift upon cooling and signals localization of charge carriers resulting in sub-diffusive transport and major deviations of the intraband optical conductivity from the simple Drude model\cite{Luca2017, Fratini2014, Fratini2020, Uykur2011}. The redshift is accompanied by a sudden reduction in the spectral weight across \Tc\ (Fig.~\ref{2}g). While the total spectral weight is conserved according to the sum rules (Fig.~\ref{2}h), the weight of the localization peak drops at \Tc, suggesting a redistribution of the intensity upon the DW formation. 

The evolution of the localization peak gives one plausible explanation for the behavior of the 482~\cm\ mode as a function of temperature. The strong asymmetry of this mode is probably caused by the coupling to the localization peak located around the same frequency. At lower temperatures, the red shift of the localization peak reduces the coupling, and the 482~\cm\ mode becomes more symmetric. Its intensity decreases (Fig.~\ref{4}h) because a fraction of the localized carriers is eliminated by the density-wave formation below \Tc.

Different microscopic scenarios of the localization peak were discussed in the literature. Whereas mere disorder in the hopping paths\cite{Fratini2014} is excluded by the high quality of our crystals, the electron-phonon\cite{Atta2004} and electron-electron\cite{Luca2017} interactions are both likely candidates. Interestingly, even at low temperatures the localization peak does not shift to zero frequency and does not merge with the Drude peak (see the inset of Fig.~\ref{2}f). This indicates that the slowing down of electron dynamics is caused not only by thermal effects, such as phonons, but also by interactions between the electrons.

Further evidence for such electronic correlations is found upon comparing the experimental optical response of \KVS\ with DFT. Experimental optical conductivity could be reproduced, but only with renormalized band energies obtained via shifting the Fermi level by $+64$\,meV. Moreover, the calculated plasma frequency of 3.48~eV is far above the experimental one of 2.12~eV. This discrepancy gives rise to a significantly overestimated Drude spectral weight, ${\rm SW}_{\rm exp}/{\rm SW}_{\rm DFT}={\omega}^2_{\rm p, exp}/{\omega}^2_{\rm p,  DFT}=0.37$ (See Supplementary Note 5 for the plasma frequency, $\omega_{\rm p}$, estimation), which is a hallmark of electronic correlations in topological semimetals according to Ref.~\onlinecite{Shao2020}. Changes in the sample stoichiometry, such as potassium deficiency, can not be responsible for this effect because they lead to a downward shift of the Fermi level, whereas the upward shift is required to reproduce the experimental optical response.


\section*{Discussion} 

The 78\,K anomaly was known since the discovery of \KVS{}, but its nature has remained enigmatic. Our broadband optical spectroscopy experiments witness a strong renormalization of the DOS across \Tc and set firm grounds for interpreting this anomaly as a charge-density-wave order. Incidentally, a prominent localization peak in the low-energy optical absorption both above and below \Tc{} signals electronic correlations that coexist with strong phonon anomalies. These findings render electron-phonon and electron-electron interactions two important ingredients of the \KVS{} physics. They also suggest that states distinct from a simple charge order may be relevant to the density wave in \KVS{}. Indeed, candidate states, which we identified for the first time, bear strong resemblance to the $p$-wave states of the kagome Hubbard model and allow the interpretation of \KVS{} as a correlated kagome metal.

Theoretical results available for the kagome Hubbard model at the van Hove singularity\cite{Kiesel2013,Wang2013} and in other regimes\cite{Ohashi2006,Guertler2013,Kudo2019,Kim2020,DiSante2020,Kaufmann2021} offer several interesting insights into the \KVS{} physics including the large extrinsic anomalous Hall effect\cite{Yang2020} in \KVS{}, as well as chiral anisotropy detected in the recent tunneling spectroscopy experiment\cite{Jiang2021}. Spin bond order (triplet $p$-wave) includes spin degrees of freedom in the form of spin currents that can not be detected via conventional probes, such as muon spectroscopy or neutron scattering\cite{Nayak2000}, but may have influence on charge transport, including anomalous Hall effect. Our tri-hexagon structure is the most stable type of a density wave in \KVS{} and displays an intriguing similarity to this theoretically predicted spin bond order. On the other hand, doping the kagome Hubbard model away from the van Hove singularity leads to superconducting instabilities\cite{Kiesel2012,Yu2012,Wang2013} that allow further interesting, yet to be explored connections to the recently established superconducting properties of \KVS{} and related materials\cite{Ortiz2020a,Yin2021,Ortiz2021,Wang2020,Du2021,Chen2021,Zhao2021,Tan2021,Liang2021}.

{\textit{Note added:} after the initial submission of this manuscript we studied the optical response of the isostructural compound \CVS\cite{Uykur2021}. Several important differences from \KVS\ are worth noting: (i) Different positions of the band saddle points at $M$ lead to a strong change in the interband absorption at low energies in \CVS\ compared to \KVS; (ii) In \CVS, this interband absorption is perfectly reproduced by DFT without any renormalization of band energies; (iii) At low temperatures, the localization peak in \CVS\ shifts to zero energy and merges with the Drude peak, unlike in \KVS\ where the peak position saturates at finite energies, as shown in the inset of Fig.~\ref{2}f. This indicates that the slowing down of electron dynamics is thermally activated in \CVS, possibly due to interactions with phonons, while in \KVS\ it is intrinsic to electrons and, therefore, persists even at low temperatures. All these observations reinforce our conclusion that electron-electron interactions must play an important role in \KVS, and that \KVS\ is significantly different from its Cs sibling. These differences may also be responsible for the complete screening of phonons in \CVS\cite{Uykur2021}, as opposed to \KVS.

\begin{methods}

\noindent\textbf{Sample Characterization.} High-quality single crystals were prepared as described elsewhere \cite{Ortiz2019, Ortiz2021}. Freshly cleaved samples with the dimensions $\sim 2\times 3 \times 0.2$~mm$^3$ were used for optical measurements. In-plane component of the optical conductivity ($\sigma_{xx}$) was probed in all measurements. dc resistivity was measured with the standard four-point contact method to control the amount of potassium on the exact same piece as used in optical experiments. Magnetic anomaly has also been confirmed via magnetic susceptibility measurements. Here, a field of 1\,T has been applied along $H\,\|\,c$ and the susceptibility has been measured in field-cooled (FC) configuration with a Quantum Design Magnetic Property Measurement System (MPMS). Comparison to the earlier literature suggests that our sample of \KVS{} is nearly stoichiometric, with the possible K deficiency level of well below 8\% reported in Ref.~\onlinecite{Ortiz2019} (see Supplementary Note 1). The optical spectra were well reproducible across measurements in different frequency ranges and on two different crystals, suggesting that no changes in the K stoichiometry could occur during the measurement and that all spectral features revealed by our study are intrinsic to \KVS.

\noindent\textbf{Optical Measurements.} Broadband reflectivity measurements were performed with two Bruker Fourier Transform Infrared (FTIR) spectrometers. For the high-energy range ($0.075-2.25$~eV / $600-18000$~\cm), an infrared microscope coupled to a VERTEX80v FTIR spectrometer is utilized, where the infrared light is focused to 200~µm$^2$. Freshly evaporated gold mirrors are used for the reference. For the low-energy measurements ($0.01-0.1$~eV / $70-700$~\cm), an IFS 113v spectrometer coupled with a custom made cryostat is used. Absolute reflectivity of the sample has been obtained with the gold overcoating technique \cite{Homes1993}. 

Optical conductivity is calculated via Kramers-Kronig (KK) analysis from the measured reflectivity. For the KK analysis, data are extrapolated using Hagen-Rubens relations to the low-energy range, while x-ray scattering functions have been utilized for the high-energy extrapolations \cite{Tanner2015}.

\noindent\textbf{Computational Details.} Density-functional (DFT) band-structure calculations were performed in the \texttt{Wien2K}\cite{wien2k,Blaha2020}, \texttt{FPLO}\cite{fplo}, and \texttt{VASP}\cite{vasp1,vasp2} codes with several cross-checks that ensured the robust nature of our computational results. In all cases, the Perdew-Burke-Ernzerhof flavor of the exchange-correlation potential\cite{pbe96} was used, and the $k$-mesh with $L\times L\times L/2$ points adapted to the anisotropic nature of the KV$_3$Sb$_5$ structure was employed (see Supplementary Note 6 for details on the choice of the $k$-mesh and convergence). Experimental structural parameters from Ref.~\onlinecite{Ortiz2019} were chosen for the undistorted \KVS{} structure, with the Sb1 atom located at $(\frac23,\frac13,0.7539)$. Note that $z_{\rm Sb1}=0.7539(1)$ is the correct value for the powder refinement of the stoichiometric \KVS{} sample in Ref.~\onlinecite{Ortiz2019}. The same value is obtained by extrapolating the results of single-crystal refinements of the K-deficient samples\cite{Ortiz2019}. Band dispersions were independently calculated in Wien2K and FPLO upon converging fully self-consistent calculations on the $k$-mesh with $L=36$. Wien2K was then used to compute optical conductivities with $L=72$. 

Crystal structures of possible DW states were relaxed in \texttt{VASP} and \texttt{FPLO} in the $2a\times 2a\times c$ supercell until residual forces were below 0.002\,eV/\r A. To facilitate comparison with the optical data and kagome Hubbard model, we focus on the charge order in the $ab$ plane and disregard possible modulation along the $c$-axis\cite{Jiang2021}. Given the doubled $a$ and $c$ parameters, these calculations were performed on the $k$-mesh with $L\times L\times L$ points and $L=16$. Consequently, optical conductivity was calculated for the fully relaxed structures in Wien2K with $L=24$. Additionally, phonon calculations were performed in \texttt{VASP} using the built-in procedure with frozen atomic displacements of 0.015\,\r A. 

All calculations were performed with spin-orbit coupling (SOC) included. Whereas the SOC has minor influence on lattice energies and phonons, it affects band structure and optical transitions in the vicinity of the Fermi level.

\noindent\textbf{Data availability.} The data that support the findings of this study are available from the corresponding authors upon request.
	
\end{methods}


\section*{Acknowledgements}
Authors acknowledge the fruitful discussions with Artem Pronin and Sascha Polatkan and technical support by Gabriele Untereiner. We also thank Berina Klis for the dc resistivity measurements. S.D.W. and B.R.O. gratefully acknowledge support via the UC Santa Barbara NSF Quantum Foundry funded via the Q-AMASE-i program under award DMR-1906325. B. R. O. also acknowledges support from the California NanoSystems Institute through the Elings fellowship program. The work has been supported by the Deutsche Forschungsgemeinschaft (DFG) via DR228/51-1 and UY62/2-1. E.U. acknowledges the European Social Fund and the Baden-W\"urttemberg Stiftung for the financial support of this research project by the Eliteprogramme. A.T. was supported by the Federal Ministry for Education and Research via the Sofja Kovalevskaya Award of Alexander von Humboldt Foundation.

\section*{Additional information}
\begin{addendum}
   	\item[Competing Interests] The Authors declare no Competing Financial or Non-Financial Interests.
    \item[Author contributions] E.U performed the experiments. A.A.T performed the DFT calculations. Samples were grown by B.R.O. and S.D.W. The manucsript was written by A.A.T, E.U., and M.D. with the suggestions from all authors.
	\item[Supplementary information] accompanies this paper at 
	\item[Correspondence] Correspondence and requests for materials should be addressed to Ece Uykur~(email: ece.uykur@pi1.physik.uni-stuttgart.de) and Alexander A. Tsirlin~(email: altsirlin@gmail.com).
\end{addendum}



\section*{References}
\begin{thebibliography}{10}
\expandafter\ifx\csname url\endcsname\relax
 \def\url#1{\texttt{#1}}\fi
\expandafter\ifx\csname urlprefix\endcsname\relax\def\urlprefix{URL }\fi
\providecommand{\bibinfo}[2]{#2}
\providecommand{\eprint}[2][]{\url{#2}}

\bibitem{Broholm2020}
\bibinfo{author}{Broholm, C.}\emph{et~al.} 
\newblock \bibinfo{title}{Quantum spin liquids}.
\newblock \emph{\bibinfo{journal}{Science}}
  \textbf{\bibinfo{volume}{367}}, \bibinfo{pages}{eaay0668}
  (\bibinfo{year}{2020}).
  
 \bibitem{Lin2018}
\bibinfo{author}{Lin, Z.} \emph{et~al.}
\newblock \bibinfo{title}{Flatbands and Emergent Ferromagnetic Ordering in {Fe$_3$Sn$_2$} Kagome Lattices}.
\newblock \emph{\bibinfo{journal}{Phys. Rev. Lett.}}
  \textbf{\bibinfo{volume}{128}}, \bibinfo{pages}{096401}
  (\bibinfo{year}{2018}).
  
 \bibitem{Ye2018}
\bibinfo{author}{Ye, L.} \emph{et~al.}
\newblock \bibinfo{title}{Massive {Dirac} fermions in a ferromagnetic kagome metal}.
\newblock \emph{\bibinfo{journal}{Nature}}
  \textbf{\bibinfo{volume}{555}}, \bibinfo{pages}{638--642}
  (\bibinfo{year}{2018}).
  
 \bibitem{Liu2019}
\bibinfo{author}{Liu, D.F.} \emph{et~al.}
\newblock \bibinfo{title}{Magnetic {Weyl} semimetal phase in a Kagom\'e crystal}.
\newblock \emph{\bibinfo{journal}{Scince}}
  \textbf{\bibinfo{volume}{365}}, \bibinfo{pages}{1282--1285}
  (\bibinfo{year}{2019}).
  
  \bibitem{Yin2019}
\bibinfo{author}{Yin, J.-X.} \emph{et~al.}
\newblock \bibinfo{title}{Negative flat band magnetism in a spin-orbit-coupled correlated kagome magnet}.
\newblock \emph{\bibinfo{journal}{Nat. Phys.}}
  \textbf{\bibinfo{volume}{15}}, \bibinfo{pages}{443--448}
  (\bibinfo{year}{2019}).
  
   \bibitem{Kang2020}
\bibinfo{author}{Kang, M.} \emph{et~al.}
\newblock \bibinfo{title}{Dirac fermions and flat bands in the ideal kagome metal {FeSn}}.
\newblock \emph{\bibinfo{journal}{Nat. Mater.}}
  \textbf{\bibinfo{volume}{19}}, \bibinfo{pages}{163--169}
  (\bibinfo{year}{2020}).
  
   \bibitem{Ye2019}
\bibinfo{author}{Ye, L.} \emph{et~al.}
\newblock \bibinfo{title}{{de Haas-van Alphen effect of correlated Dirac states in kagome metal Fe$_3$Sn$_2$}}.
\newblock \emph{\bibinfo{journal}{Nat. Commun.}}
  \textbf{\bibinfo{volume}{10}}, \bibinfo{pages}{4870}
  (\bibinfo{year}{2019}).
  
   \bibitem{Guguchia2020}
\bibinfo{author}{Guguchia, Z.} \emph{et~al.}
\newblock \bibinfo{title}{Tunable anomalous {Hall} conductivity through volume-wise magnetic competition in a topological kagome magnet}.
\newblock \emph{\bibinfo{journal}{Nat. Comm.}}
  \textbf{\bibinfo{volume}{11}}, \bibinfo{pages}{559}
  (\bibinfo{year}{2020}).
  
   \bibitem{Biswas2020}
\bibinfo{author}{Biswas, A.} \emph{et~al.}
\newblock \bibinfo{title}{{Spin-Reorientation-Induced Band Gap in ${\mathrm{Fe}}_{3}{\mathrm{Sn}}_{2}$: Optical Signatures of Weyl Nodes}}.
\newblock \emph{\bibinfo{journal}{Phys. Rev. Lett.}}
  \textbf{\bibinfo{volume}{125}}, \bibinfo{pages}{076403}
  (\bibinfo{year}{2020}).
  
   \bibitem{Ortiz2019}
\bibinfo{author}{Ortiz, B.R.} \emph{et~al.}
\newblock \bibinfo{title}{{New kagome prototype materials: discovery of KV$_3$Sb$_5$, RbV$_3$Sb$_5$, and CsV$_3$Sb$_5$}}.
\newblock \emph{\bibinfo{journal}{Phys. Rev. Mater.}}
  \textbf{\bibinfo{volume}{3}}, \bibinfo{pages}{094407}
  (\bibinfo{year}{2019}).
  
   \bibitem{Yin2021}
\bibinfo{author}{Yin, Q.} \emph{et~al.}
\newblock \bibinfo{title}{{Superconductivity and Normal-State Properties of Kagome Metal RbV$_3$Sb$_5$ Single Crystals}}.
\newblock \emph{\bibinfo{journal}{Chin. Phys. Lett.}}
  \textbf{\bibinfo{volume}{38}}, \bibinfo{pages}{037403}
  (\bibinfo{year}{2021}).
  
   \bibitem{Ortiz2020a}
\bibinfo{author}{Ortiz, B.R.} \emph{et~al.}
\newblock \bibinfo{title}{sV$_3$Sb$_5$: A $Z_{2}$ Topological Kagome Metal with a Superconducting Ground State}.
\newblock \emph{\bibinfo{journal}{Phys. Rev. Lett.}}
  \textbf{\bibinfo{volume}{125}}, \bibinfo{pages}{247002}
  (\bibinfo{year}{2020}).
  
 \bibitem{Ortiz2021}
\bibinfo{author}{Ortiz, B.R.} \emph{et~al.}
\newblock \bibinfo{title}{Superconductivity in the {$Z_{2}$ kagome metal KV$_3$Sb$_5$}}.
\newblock \emph{\bibinfo{journal}{Phys. Rev. Mater.}}
  \textbf{\bibinfo{volume}{5}}, \bibinfo{pages}{034801}
  (\bibinfo{year}{2021}).
  
   \bibitem{Kenney2021}
\bibinfo{author}{Kenney, E. M., Ortiz, B. R., Wang, C., Wilson, S. D. \& Graf, M. J.}
\newblock \bibinfo{title}{{Absence of local moments in the kagome metal KV$_3$Sb$_5$ as determined by muon spin spectroscopy}}.
\newblock \emph{\bibinfo{journal}{J. Phys. Condens. Matter}}
  \textbf{\bibinfo{volume}{33}}, \bibinfo{pages}{235801}
  (\bibinfo{year}{2021}).
  
   \bibitem{Yang2020}
\bibinfo{author}{Yang, S.-Y.} \emph{et~al.}
\newblock \bibinfo{title}{Giant, unconventional anomalous {Hall} effect in the metallic frustrated magnet candidate, {KV$_3$Sb$_5$}}.
\newblock \emph{\bibinfo{journal}{Sci. Adv.}}
  \textbf{\bibinfo{volume}{6}}, \bibinfo{pages}{eabb6003}
  (\bibinfo{year}{2020}).
  
   \bibitem{Yu2021}
\bibinfo{author}{Yu, F. H.} \emph{et~al.}
\newblock \bibinfo{title}{{Concurrence of anomalous Hall effect and charge density wave in a superconducting topological kagome metal}}.
\newblock \emph{\bibinfo{journal}{Phys. Rev. B}}
  \textbf{\bibinfo{volume}{104}}, \bibinfo{pages}{1041103}
  (\bibinfo{year}{}).
  
   \bibitem{Jiang2021}
\bibinfo{author}{Jiang, Y.-X.} \emph{et~al.}
\newblock \bibinfo{title}{{Unconventional chiral charge order in kagome superconductor KV$_3$Sb$_5$}}.
\newblock \emph{\bibinfo{journal}{Nat. Mater.}}
  \textbf{\bibinfo{volume}{20}}, \bibinfo{pages}{1353–1357}
  (\bibinfo{year}{2021}).
  
   \bibitem{Kiesel2012}
\bibinfo{author}{Kiesel, M. L. \& Thomale, R.}
\newblock \bibinfo{title}{Sublattice interference in the kagome {Hubbard} model}.
\newblock \emph{\bibinfo{journal}{Phys. Rev. B}}
  \textbf{\bibinfo{volume}{86}}, \bibinfo{pages}{121105(R)}
  (\bibinfo{year}{2012}).
  
   \bibitem{Kiesel2013}
\bibinfo{author}{Kiesel, M. L., Platt, C. \& Thomale, R.} 
\newblock \bibinfo{title}{Unconventional {Fermi} Surface Instabilities in the Kagome {Hubbard} Model}.
\newblock \emph{\bibinfo{journal}{Phys. Rev. Lett.}}
  \textbf{\bibinfo{volume}{110}}, \bibinfo{pages}{126405}
  (\bibinfo{year}{2013}).
  
   \bibitem{Nayak2000}
\bibinfo{author}{Nayak, C.} 
\newblock \bibinfo{title}{Density-wave states of nonzero angular momentum}.
\newblock \emph{\bibinfo{journal}{Phys. Rev. B}}
  \textbf{\bibinfo{volume}{62}}, \bibinfo{pages}{4880}
  (\bibinfo{year}{2000}).
  
   \bibitem{Johannes2008}
\bibinfo{author}{Johannes, M. D. \& Mazin, I. I.} 
\newblock \bibinfo{title}{{Fermi surface nesting and the origin of charge density waves in metals}}.
\newblock \emph{\bibinfo{journal}{Phys. Rev. B}}
  \textbf{\bibinfo{volume}{77}}, \bibinfo{pages}{165135}
  (\bibinfo{year}{2008}).
  
   \bibitem{Wang2013}
\bibinfo{author}{Wang, W.-S., Li, Z.-Z.,  Xiang, Y.-Y. \& Wang, Q.-H.} 
\newblock \bibinfo{title}{ompeting electronic orders on kagome lattices at van {Hove} filling}.
\newblock \emph{\bibinfo{journal}{Phys. Rev. B}}
  \textbf{\bibinfo{volume}{87}}, \bibinfo{pages}{115135}
  (\bibinfo{year}{2013}).
  
   \bibitem{Gusynin2006}
\bibinfo{author}{Gusynin, V. P., Sharapov, S. G. \& Carbotte, J. P.} 
\newblock \bibinfo{title}{Unusual Microwave Response of Dirac Quasiparticles in Graphene}.
\newblock \emph{\bibinfo{journal}{Phys. Rev. Lett.}}
  \textbf{\bibinfo{volume}{96}}, \bibinfo{pages}{256802}
  (\bibinfo{year}{2006}).
  
   \bibitem{Kuzmenko2008}
\bibinfo{author}{Kuzmenko, A. B., van Heumen, E., Carbone, F. \& van der Marel, D.}
\newblock \bibinfo{title}{Universal Optical Conductance of Graphite}.
\newblock \emph{\bibinfo{journal}{Phys. Rev. Lett.}}
  \textbf{\bibinfo{volume}{100}}, \bibinfo{pages}{117401}
  (\bibinfo{year}{2008}).
  
   \bibitem{Zhu2015}
\bibinfo{author}{Zhu, X., Cao, Y., Zhang, J., Plummer, E. W. \& Guo, J.} 
\newblock \bibinfo{title}{Classification of charge density waves based on their nature}.
\newblock \emph{\bibinfo{journal}{Proc. Nat. Acad. Sci.}}
  \textbf{\bibinfo{volume}{112}}, \bibinfo{pages}{2367--2371}
  (\bibinfo{year}{}).
  
   \bibitem{Basov2011}
\bibinfo{author}{Basov, D. N., Averitt, R. D., van der Marel, D., Dressel, M. \& Haule, K.} 
\newblock \bibinfo{title}{Electrodynamics of correlated electron materials}.
\newblock \emph{\bibinfo{journal}{Rev. Mod. Phys.}}
  \textbf{\bibinfo{volume}{83}}, \bibinfo{pages}{471--541}
  (\bibinfo{year}{2011}).
    
    
 \bibitem{Zhao2021}
\bibinfo{author}{Zhao, C.C.} \emph{et~al.}
\newblock \bibinfo{title}{Nodal superconductivity and superconducting dome in the topological Kagome metal {CsV$_3$Sb$_5$}}.
\newblock \bibinfo{note}{Preprint at \href{https://arxiv.org/abs/2102.08356}{https://arxiv.org/abs/2102.08356}}
  (\bibinfo{year}{2021}).
  
   \bibitem{Wilson1975}
\bibinfo{author}{Wilson, J. A., Di Salvo, F. J. \& Mahajan, S. } 
\newblock \bibinfo{title}{Charge-density waves and superlattices in the metallic layered transition metal dichalcogenides}.
\newblock \emph{\bibinfo{journal}{Adv. Phys.}}
  \textbf{\bibinfo{volume}{24}}, \bibinfo{pages}{117-201}
  (\bibinfo{year}{1975}).
  
   \bibitem{Fazekas1979}
\bibinfo{author}{Fazekas, E. \& Tosatti, E.} 
\newblock \bibinfo{title}{Electrical, structural and magnetic properties of pure and doped {1T-TaS$_2$}}.
\newblock \emph{\bibinfo{journal}{Philos. Mag. B}}
  \textbf{\bibinfo{volume}{39}}, \bibinfo{pages}{229-244}
  (\bibinfo{year}{1979}).
  
   \bibitem{Wang2020a}
\bibinfo{author}{Wang, Y. D.} \emph{et~al.}
\newblock \bibinfo{title}{Band insulator to {Mott} insulator transition in {1T-TaS$_2$}}.
\newblock \emph{\bibinfo{journal}{Nat. Commun.}}
  \textbf{\bibinfo{volume}{11}}, \bibinfo{pages}{4215}
  (\bibinfo{year}{2020}).
  
   \bibitem{Comin2014}
\bibinfo{author}{Comin, R.} \emph{et~al.}
\newblock \bibinfo{title}{Charge order driven by Fermi-arc instability in Bi$_2$Sr$_{2-x}$La$_x$CuO$_{6+\delta}$}.
\newblock \emph{\bibinfo{journal}{Science}}
  \textbf{\bibinfo{volume}{343}}, \bibinfo{pages}{390--392}
  (\bibinfo{year}{2014}).
  
   \bibitem{Fujita2014}
\bibinfo{author}{Fujita, K.} \emph{et~al.}
\newblock \bibinfo{title}{{Direct phase-sensitive identification of a $d$-form factor density wave in underdoped cuprates}}.
\newblock \emph{\bibinfo{journal}{Proc. Nat. Acad. Sci.}}
  \textbf{\bibinfo{volume}{111}}, \bibinfo{pages}{E3026--E3032}
  (\bibinfo{year}{2014}).
  
   \bibitem{Du2013}
\bibinfo{author}{Du, L., Mackeprang, K. \& Kjaergaard, H. G.} 
\newblock \bibinfo{title}{Fundamental and overtone vibrational spectroscopy{,} enthalpy of hydrogen bond formation and equilibrium constant determination of the methanol–dimethylamine complex}.
\newblock \emph{\bibinfo{journal}{Phys. Chem. Chem. Phys.}}
  \textbf{\bibinfo{volume}{15}}, \bibinfo{pages}{10194-10206}
  (\bibinfo{year}{2013}).
  
   \bibitem{Lapointe2017}
\bibinfo{author}{Lapointe, F.} \emph{et~al.}
\newblock \bibinfo{title}{{Antiresonances in the mid-infrared vibrational spectrum of functionalized graphene}}.
\newblock \emph{\bibinfo{journal}{J. Phys. Chem. C}}
  \textbf{\bibinfo{volume}{121}}, \bibinfo{pages}{9053--9062}
  (\bibinfo{year}{2017}).
  
   \bibitem{Tan2021}
\bibinfo{author}{Tan, H., Liu, Y., Wang, Z. \& Yan, Binghai} 
\newblock \bibinfo{title}{Charge Density Waves and Electronic Properties of Superconducting Kagome Metals}.
\newblock \emph{\bibinfo{journal}{Phys. Rev. Lett.}}
  \textbf{\bibinfo{volume}{127}}, \bibinfo{pages}{046401}
  (\bibinfo{year}{2021}).
  
   \bibitem{Bonini2007}
\bibinfo{author}{Bonini, N., Lazzeri, M., Marzari, N. \& Mauri, F.} 
\newblock \bibinfo{title}{Phonon Anharmonicities in Graphite and Graphene}.
\newblock \emph{\bibinfo{journal}{Phys. Rev. Lett.}}
  \textbf{\bibinfo{volume}{99}}, \bibinfo{pages}{176802}
  (\bibinfo{year}{2007}).
  
   \bibitem{Lazzeri2006}
\bibinfo{author}{Lazzeri, M., Piscanec, S., Mauri, F., Ferrari, A. C. \& Robertson, J.} 
\newblock \bibinfo{title}{Phonon linewidths and electron-phonon coupling in graphite and nanotubes}.
\newblock \emph{\bibinfo{journal}{Phys. Rev. B}}
  \textbf{\bibinfo{volume}{73}}, \bibinfo{pages}{155426}
  (\bibinfo{year}{2006}).
  
   \bibitem{Luca2017}
\bibinfo{author}{Luca, V. D., Gout\'{e}aux, B., Hartnoll, S. A. \& Karlsson, A.} 	
\newblock \bibinfo{title}{Bad metals from fluctuating density waves}.
\newblock \emph{\bibinfo{journal}{SciPost Phys.}}
  \textbf{\bibinfo{volume}{3}}, \bibinfo{pages}{025}
  (\bibinfo{year}{2017}).
  
   \bibitem{Monceau2012}
\bibinfo{author}{Monceau, P.} 
\newblock \bibinfo{title}{Electronic crystals: An experimental overview}.
\newblock \emph{\bibinfo{journal}{Adv. Phys.}}
  \textbf{\bibinfo{volume}{61}}, \bibinfo{pages}{325-581}
  (\bibinfo{year}{2012}).
    
    
 \bibitem{Calandra2009}
\bibinfo{author}{Calandra, M., Mazin, I. I. \& Mauri, F.}
\newblock \bibinfo{title}{{Effect of dimensionality on the charge-density wave in few-layer $2{H\text{-NbSe}}_{2}$}}.
\newblock \emph{\bibinfo{journal}{Phys. Rev. B	}}
  \textbf{\bibinfo{volume}{80}}, \bibinfo{pages}{241108}
  (\bibinfo{year}{2009}).
  
     \bibitem{Fratini2014}
\bibinfo{author}{Fratini, S. and Ciuchi, S. \& Mayou, D.} 
\newblock \bibinfo{title}{{Phenomenological model for charge dynamics and optical response of disordered systems: Application to organic semiconductors}}.
\newblock \emph{\bibinfo{journal}{Phys. Rev. B}}
  \textbf{\bibinfo{volume}{89}}, \bibinfo{pages}{235201}
  (\bibinfo{year}{2014}).
  
   \bibitem{Fratini2020}
\bibinfo{author}{Fratini, S. \& Ciuchi, S.} 
\newblock \bibinfo{title}{Dynamical localization corrections to band transport}.
\newblock \emph{\bibinfo{journal}{Phys. Rev. Res.}}
  \textbf{\bibinfo{volume}{2}}, \bibinfo{pages}{013001}
  (\bibinfo{year}{2020}).
  
   \bibitem{Uykur2011}
\bibinfo{author}{Uykur, E. and Tanaka, K. and Masui, T. and Miyasaka, S. \& Tajima, S.} 
\newblock \bibinfo{title}{{In-plane optical spectra of Y${}_{1\ensuremath{-}x}$Ca${}_{x}$Ba${}_{2}$Cu${}_{3}$O${}_{7\ensuremath{-}\ensuremath{\delta}}$: Overdoping and disorder effects on residual conductivity}}.
\newblock \emph{\bibinfo{journal}{Phys. Rev. B}}
  \textbf{\bibinfo{volume}{84}}, \bibinfo{pages}{184527}
  (\bibinfo{year}{2011}).
  
   \bibitem{Uykur2021}
\bibinfo{author}{Uykur, E.} \emph{et~al.}
\newblock \bibinfo{title}{{Low-energy optical properties of the nonmagnetic kagome metal CsV$_3$Sb$_5$}}.
\newblock \emph{\bibinfo{journal}{Phys. Rev. B}}
  \textbf{\bibinfo{volume}{104}}, \bibinfo{pages}{045130}
  (\bibinfo{year}{2021}).
  
   \bibitem{Atta2004}
\bibinfo{author}{Atta-Fynn, R., Biswas, P. \& Drabold, D. A.} 
\newblock \bibinfo{title}{{Electron--phonon coupling is large for localized states}}.
\newblock \emph{\bibinfo{journal}{Phys. Rev. B}}
  \textbf{\bibinfo{volume}{69}}, \bibinfo{pages}{245204}
  (\bibinfo{year}{2004}).
  
   \bibitem{Ohashi2006}
\bibinfo{author}{Ohashi, T., Kawakami, N. \& Tsunetsugu, Hirokazu} 
\newblock \bibinfo{title}{Mott Transition in Kagom\'e Lattice {Hubbard} Model}.
\newblock \emph{\bibinfo{journal}{Phys. Rev. Lett.}}
  \textbf{\bibinfo{volume}{97}}, \bibinfo{pages}{066401}
  (\bibinfo{year}{2006}).
  
   \bibitem{Shao2020}
\bibinfo{author}{Shao, Y.} \emph{et~al.}
\newblock \bibinfo{title}{Electronic correlations in nodal-line semimetals}.
\newblock \emph{\bibinfo{journal}{Nat. Phys.}}
  \textbf{\bibinfo{volume}{16}}, \bibinfo{pages}{636-641}
  (\bibinfo{year}{2020}).
  
   \bibitem{Guertler2013}
\bibinfo{author}{Guertler, S. \& Monien, H.} 
\newblock \bibinfo{title}{Unveiling the Physics of the Doped Phase of the {$t-J$} Model on the Kagome Lattice}.
\newblock \emph{\bibinfo{journal}{Phys. Rev. Lett.}}
  \textbf{\bibinfo{volume}{111}}, \bibinfo{pages}{097204}
  (\bibinfo{year}{2013}).
  
   \bibitem{Kudo2019}
\bibinfo{author}{Kudo, K., Yoshida, T. \& Hatsugai, Yasuhiro} 
\newblock \bibinfo{title}{Higher-Order Topological {Mott} Insulators}.
\newblock \emph{\bibinfo{journal}{Phys. Rev. Lett.}}
  \textbf{\bibinfo{volume}{123}}, \bibinfo{pages}{196402}
  (\bibinfo{year}{2019}).
  
   \bibitem{DiSante2020}
\bibinfo{author}{{Di Sante}, D.} \emph{et~al.}
\newblock \bibinfo{title}{Turbulent hydrodynamics in strongly correlated Kagome metals}.
\newblock \emph{\bibinfo{journal}{Nat. Commun.}}
  \textbf{\bibinfo{volume}{11}}, \bibinfo{pages}{3997}
  (\bibinfo{year}{2020}).
  
   \bibitem{Kaufmann2021}
\bibinfo{author}{Kaufmann, J., Steiner, K., Scalettar, R. T., Held, K. \& Janson, O.} 
\newblock \bibinfo{title}{How correlations change the magnetic structure factor of the kagome Hubbard model}.
\newblock \emph{\bibinfo{journal}{Phys. Rev. B}}
  \textbf{\bibinfo{volume}{104}}, \bibinfo{pages}{165127}
  (\bibinfo{year}{2021}).
  
   \bibitem{Kim2020}
\bibinfo{author}{im, H. S., Mishra, A. \& Lee, S.} 
\newblock \bibinfo{title}{Emergent chiral spin ordering and anomalous {Hall} effect in a kagome lattice at a $\frac13$ filling}.
\newblock \emph{\bibinfo{journal}{Phys. Rev. B}}
  \textbf{\bibinfo{volume}{102}}, \bibinfo{pages}{155113}
  (\bibinfo{year}{2020}).
  
   \bibitem{Yu2012}
\bibinfo{author}{Yu, S.-L. \& Li, J.-X.} 
\newblock \bibinfo{title}{Chiral superconducting phase and chiral spin-density-wave phase in a {Hubbard} model on the kagome lattice}.
\newblock \emph{\bibinfo{journal}{Phys. Rev. B}}
  \textbf{\bibinfo{volume}{85}}, \bibinfo{pages}{144402}
  (\bibinfo{year}{2012}).
    
    
 \bibitem{Wang2020}
\bibinfo{author}{Wang Y.} \emph{et~al.}
\newblock \bibinfo{title}{Proximity-induced spin-triplet superconductivity and edge supercurrent in the topological Kagome metal, {K$_{1-x}$V$_3$Sb$_5$}}.
\newblock \bibinfo{note}{Preprint at \href{https://arxiv.org/abs/2012.05898}{https://arxiv.org/abs/2012.05898}}
   (\bibinfo{year}{2020}).
  
     \bibitem{Du2021}
\bibinfo{author}{Du, F.} \emph{et~al.}
\newblock \bibinfo{title}{{Pressure-induced double superconducting domes and charge instability in the kagome metal KV$_3$Sb$_5$}}.
\newblock \emph{\bibinfo{journal}{Phys. Rev. B}}
  \textbf{\bibinfo{volume}{103}}, \bibinfo{pages}{1220504}
  (\bibinfo{year}{2021}).
  
   \bibitem{Chen2021}
\bibinfo{author}{Chen, K. Y.} \emph{et~al.}
\newblock \bibinfo{title}{Double Superconducting Dome and Triple Enhancement of T$_c$ in the Kagome Superconductor {CsV$_3$Sb$_5$} under High Pressure}.
\newblock \emph{\bibinfo{journal}{Phys. Rev. Lett.}}
  \textbf{\bibinfo{volume}{126}}, \bibinfo{pages}{247001}
  (\bibinfo{year}{2021}).
  
   \bibitem{Liang2021}
\bibinfo{author}{Liang, Z.} \emph{et~al.}
\newblock \bibinfo{title}{Three-Dimensional Charge Density Wave and Surface-Dependent Vortex-Core States in a Kagome Superconductor {CsV$_3$Sb$_5$}}.
\newblock \emph{\bibinfo{journal}{Phys. Rev. X}}
  \textbf{\bibinfo{volume}{11}}, \bibinfo{pages}{031026}
  (\bibinfo{year}{2021}).
  
   \bibitem{Homes1993}
\bibinfo{author}{Homes, C. C., Reedyk, M., Cradles, D. A. \& Timusk, T.}
\newblock \bibinfo{title}{Technique for measuring the reflectance of irregular, submillimeter-sized samples}.
\newblock \emph{\bibinfo{journal}{Appl. Opt.}}
  \textbf{\bibinfo{volume}{32}}, \bibinfo{pages}{2976--2983}
  (\bibinfo{year}{1993}).
  
   \bibitem{Tanner2015}
\bibinfo{author}{Tanner, D. B.}
\newblock \bibinfo{title}{Use of x-ray scattering functions in Kramers-Kronig analysis of reflectance}.
\newblock \emph{\bibinfo{journal}{Phys. Rev. B}}
  \textbf{\bibinfo{volume}{91}}, \bibinfo{pages}{035123}
  (\bibinfo{year}{2015}).
  
   \bibitem{Blaha2020}
\bibinfo{author}{Blaha, P.} \emph{et~al.}
\newblock \bibinfo{title}{{WIEN2k: An APW+lo program for calculating the properties of solids}}.
\newblock \emph{\bibinfo{journal}{J. Chem. Phys.}}
  \textbf{\bibinfo{volume}{152}}, \bibinfo{pages}{074101}
  (\bibinfo{year}{2020}).
  
   \bibitem{wien2k}
\bibinfo{author}{Blaha, P.} \emph{et~al.}
\newblock \bibinfo{title}{WIEN2k, an augmented plane wave + local orbitals program for calculating crystal properties (Karlheinz Schwarz, Techn. Universit\"at Wien, Austria), 2018. ISBN 3-9501031-1-2}.
  
   \bibitem{fplo}
\bibinfo{author}{Koepernik, K. \& Eschrig, H.} 
\newblock \bibinfo{title}{Full-potential nonorthogonal local-orbital minimum-basis band-structure scheme}.
\newblock \emph{\bibinfo{journal}{Phys. Rev. B}}
  \textbf{\bibinfo{volume}{59}}, \bibinfo{pages}{1743}
  (\bibinfo{year}{1999}).
  
   \bibitem{pbe96}
\bibinfo{author}{Perdew, J. P. and Burke, K. \& Ernzerhof, M.} 
\newblock \bibinfo{title}{Generalized Gradient Approximation Made Simple}.
\newblock \emph{\bibinfo{journal}{Phys. Rev. Lett.}}
  \textbf{\bibinfo{volume}{77}}, \bibinfo{pages}{3865}
  (\bibinfo{year}{1996}).
  
   \bibitem{vasp1}
\bibinfo{author}{Kresse, G. \& Furthm\"uller, J.} 
\newblock \bibinfo{title}{Efficiency of \textit{ab-initio} total energy calculations for metals and semiconductors using a plane-wave basis set}.
\newblock \emph{\bibinfo{journal}{Comput. Mater. Sci.}}
  \textbf{\bibinfo{volume}{6}}, \bibinfo{pages}{15}
  (\bibinfo{year}{1996}).
  
   \bibitem{vasp2}
\bibinfo{author}{Kresse, G. \& Furthm\"uller, J.} 
\newblock \bibinfo{title}{Efficient iterative schemes for \textit{ab initio} total-energy calculations using a plane-wave basis set}.
\newblock \emph{\bibinfo{journal}{Phys. Rev. B}}
  \textbf{\bibinfo{volume}{54}}, \bibinfo{pages}{11169}
  (\bibinfo{year}{1996}).
  
   \bibitem{Grigorenko2012}
\bibinfo{author}{Grigorenko, A. N. and Polini, M. \& Novoselov, K. S.} 
\newblock \bibinfo{title}{Graphene plasmonics}.
\newblock \emph{\bibinfo{journal}{Nat. Photonics}}
  \textbf{\bibinfo{volume}{6}}, \bibinfo{pages}{749--758}
  (\bibinfo{year}{2012}).
 
     
\end{thebibliography}

\begin{thebibliography}{10}
\expandafter\ifx\csname url\endcsname\relax
 \def\url#1{\texttt{#1}}\fi
\expandafter\ifx\csname urlprefix\endcsname\relax\def\urlprefix{URL }\fi
\providecommand{\bibinfo}[2]{#2}
\providecommand{\eprint}[2][]{\url{#2}}

\bibitem{Yang2020}
\bibinfo{author}{Yang, S.-Y.} \emph{et~al.}
\newblock \bibinfo{title}{Giant, unconventional anomalous {Hall} effect in the metallic frustrated magnet candidate, {KV$_3$Sb$_5$}}.
\newblock \emph{\bibinfo{journal}{Sci. Adv.}}
  \textbf{\bibinfo{volume}{6}}, \bibinfo{pages}{eabb6003}
  (\bibinfo{year}{2020}).
  
    \bibitem{Jiang2021}
\bibinfo{author}{Jiang, Y.-X.} \emph{et~al.}
\newblock \bibinfo{title}{{Unconventional chiral charge order in kagome superconductor KV$_3$Sb$_5$}}.
\newblock \emph{\bibinfo{journal}{Nat. Mater.}}
  \textbf{\bibinfo{volume}{20}}, \bibinfo{pages}{1353–1357}
  (\bibinfo{year}{2021}).

\end{thebibliography}

\begin{figure}
	\centering
	\includegraphics[width=1\columnwidth]{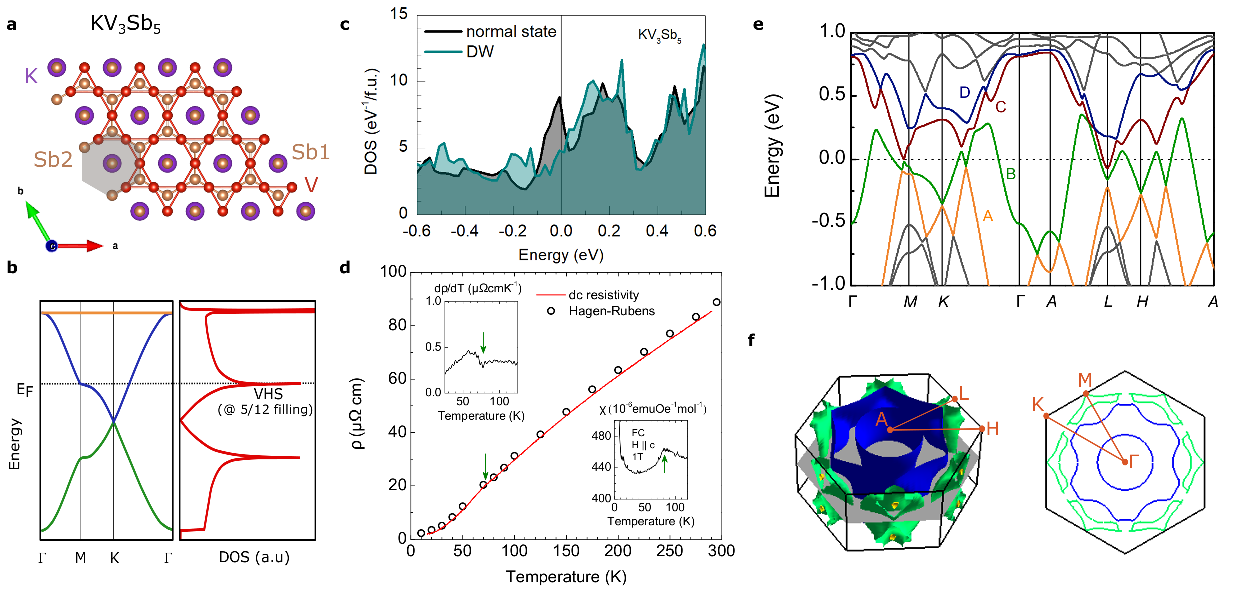}
	\caption{{\bf Crystal and band structure and the 80~K anomaly of \KVS.} {\bf a} Crystal structure of \KVS\ consists of V-Sb kagome layers and graphite-like Sb layers separated by the potassium atoms. {\bf b} Representative band structure and corresponding density-of-states (DOS) for the nearest-neighbor tight-binding model on the kagome lattice. Fermi energy is set to the van Hove singularity point at the filling factor of 5/12. {\bf c} Calculated DOS of \KVS{} in its normal (undistorted) structure and in the DW state (star structure, see Fig.~\ref{3}) demonstrates a depression in the energy spectrum at the Fermi level (E$_F$ = 0) below the transition. {\bf d} The dc resistivity of the sample used in the optical study is shown together with the resistivity values obtained by the Hagen-Rubens fits of the optical data. The insets display the 78\,K anomaly in the first derivative of the resistivity and in the magnetic susceptibility. {\bf e} The calculated band structure of \KVS\ (normal state) demonstrates the linearly dispersing bands around the Fermi energy. Relevant bands for the optical transitions have been labelled bottom up. {\bf f} Fermi surface constructed from the calculated band structure and the cut along $k_z=0$ for \KVS. The high-symmetry points and the $k$-path are shown in orange. }		
	\label{1}
\end{figure}


\begin{figure}
	\centering
	\includegraphics[width=1\columnwidth]{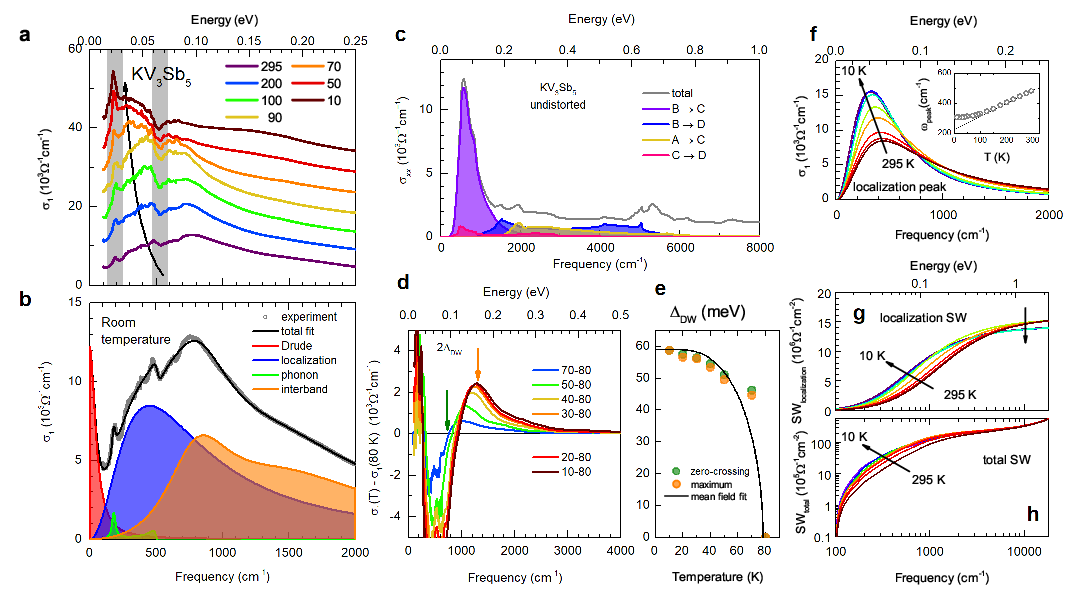}
	\caption{{\bf Optical conductivity and the density-wave state.} {\bf a} Temperature-dependent optical conductivity. Spectra shifted by 5000~\cm\ for clarity. Phonon modes are highlighted and the localization peak is demonstrated with the arrow. Optical conductivity in whole energy range is given in the Supplementary Note 2. {\bf b} Decomposition of the room-temperature optical conductivity: the low-energy Drude-contribution,  low-energy localization peak, phonon modes, and interband transitions are visible. {\bf c} Band-resolved optical conductivity for the undistorted crystal structure with the Fermi level $+64$\,meV (see Supplementary Note 6 for details). {\bf d} Difference optical conductivity in the DW state. A spectral weight transfer from low to high energies is observed. Arrows indicate the zero crossing and the maximum of the transferred peak. {\bf e} Temperature evolution of the spectral weight transfer, where the peak position is normalized to the 10~K point. A clear gap opening below 80~K is marked that also deviated from the mean-field behavior (solid line). {\bf f} Temperature evolution of the localization peak with the visible red shift upon cooling. The inset shows the peak position as a function of temperature, where a saturation at low temperatures is seen. {\bf g} SW of the localization peak showing an abrupt drop across \Tc. {\bf h} Overall SW as a function of temperature and frequency. The SW is conserved within the measured energy range. Details of the SW analysis is given in Supplementary Note 4. }		
	\label{2}
\end{figure}


\begin{figure}
	\centering
	\includegraphics[width=1\columnwidth]{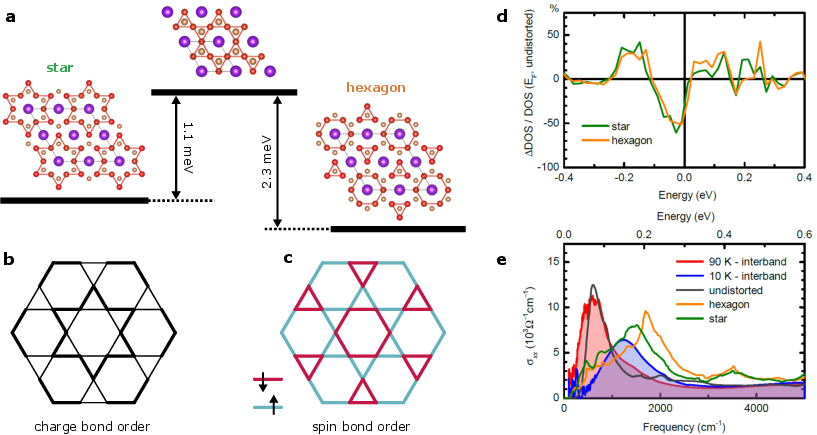}
	\caption{{\bf Density-wave states in \KVS.} {\bf a} Energy diagram of the possible DW structures in \KVS{}. The energies are given per formula unit, and the bonds show V--V distances shorter than 2.71\,\r A in the DW structures. {\bf b,c}  $p$-wave states predicted for the kagome Hubbard model at the van Hove singularity \cite{Kiesel2013}. Charge bond order in {\bf b} features spatial modulation of charge, whereas spin bond order in {\bf c}  features spatial modulation of spin current. {\bf d} Difference DOS of the undistorted and DW structures demonstrates the redistribution of states across \Tc\ on the 0.1\,eV energy scale, which is compatible with \DCDW\ determined experimentally. {\bf e}  Comparison of the experimental interband transitions with the DFT calculations in the normal and DW states.}		
	\label{3}
\end{figure}


\begin{figure}
	\centering
	\includegraphics[width=1\columnwidth]{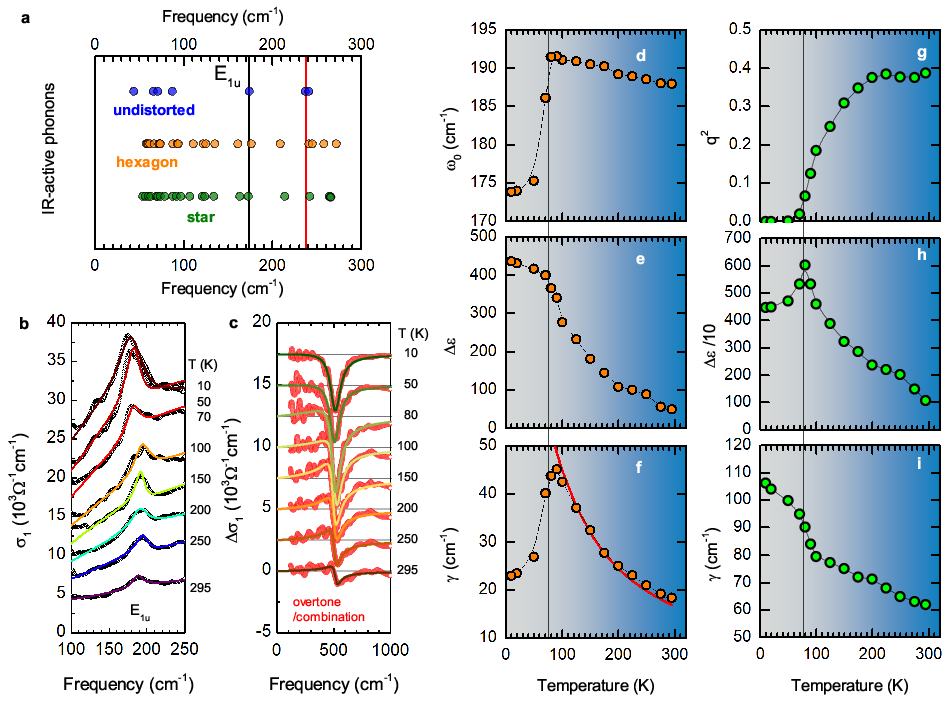}
	\caption{{\bf Phonon modes in \KVS.} {\bf a} The calculated IR-active modes are given for all structures. Vertical lines represent the phonon modes relevant to the experimental observations. The black line shows the low-energy phonon mode around 188~\cm. The red line indicates the phonon modes that can create the overtone/combination high-energy mode around 482~\cm in the IR spectra. {\bf b} Fits to the 188\,\cm\ E$_{1u}$ mode for the selected temperatures. The spectra are shifted by 2000~\cm\ for clarity. {\bf c} Fits to the 482\,\cm\ mode for the selected temperatures. The spectra are represented as $\Delta\sigma_1(\omega)$ with all other contributions subtracted and the baselines shifted by 2500 \cm\ for clarity. {\bf d-f} Fit parameters obtained from the curves in {\bf b}. The red line in {\bf f} is the expected broadening behavior due to the electron-phonon coupling \cite{Bonini2007}. {\bf g-i} Fit parameters obtained from the curves in {\bf c}. The solid lines mark \Tc.}		
	\label{4}
\end{figure}


\newpage
\begin{center}
\textbf{Supplementary information for\\ ``Optical detection of the density-wave instability in the kagome metal KV$_3$Sb$_5$"}
\end{center}
\centerline{Ece Uykur, Brenden R. Ortiz, Stephen D. Wilson, Martin Dressel, Alexander A. Tsirlin}
\newenvironment{modified}{\par\color{red}}{\par}

\renewcommand{\figurename}{Supplementary Figure}
\setcounter{figure}{0}

\textbf{This PDF file includes:\\} 
Supplementary Note 1. Comparison of the K-content\\
Supplementary Note 2. Frequency-dependent reflectivity and optical conductivity\\
Supplementary Note 3. Fit of the spectra and phonon modes\\
Supplementary Note 4. Spectral weight analysis\\
Supplementary Note 5. Plasma frequency\\
Supplementary Note 6. Additional computational results\\
Supplementary Figures 1-6\\
Supplementary References

\section*{Supplementary Note 1. Comparison of the K-content}
\KVS\ samples are prone to be K-deficient, and intentional deintercalation of K atoms can even be used to tune the Fermi energy in these materials. We compared the sample used in our optical measurements with the ones from the literature, where the K-content was known. The comparison in Supplementary Fig.~\ref{dccomparison} suggests that the sample used in the optical measurements is nearly stoichiometric.

\begin{figure}
	\centering
	\includegraphics[width=0.5\columnwidth]{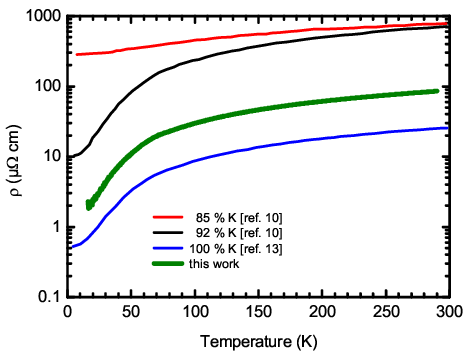}
	\caption{Dc resistivity comparison of different \KVS\ samples from the literature with the one used in the optical measurements.}		
	\label{dccomparison}
\end{figure}

\section*{Supplementary Note 2. Frequency-dependent reflectivity and optical conductivity}

\begin{figure}
	\centering
	\includegraphics[width=1\columnwidth]{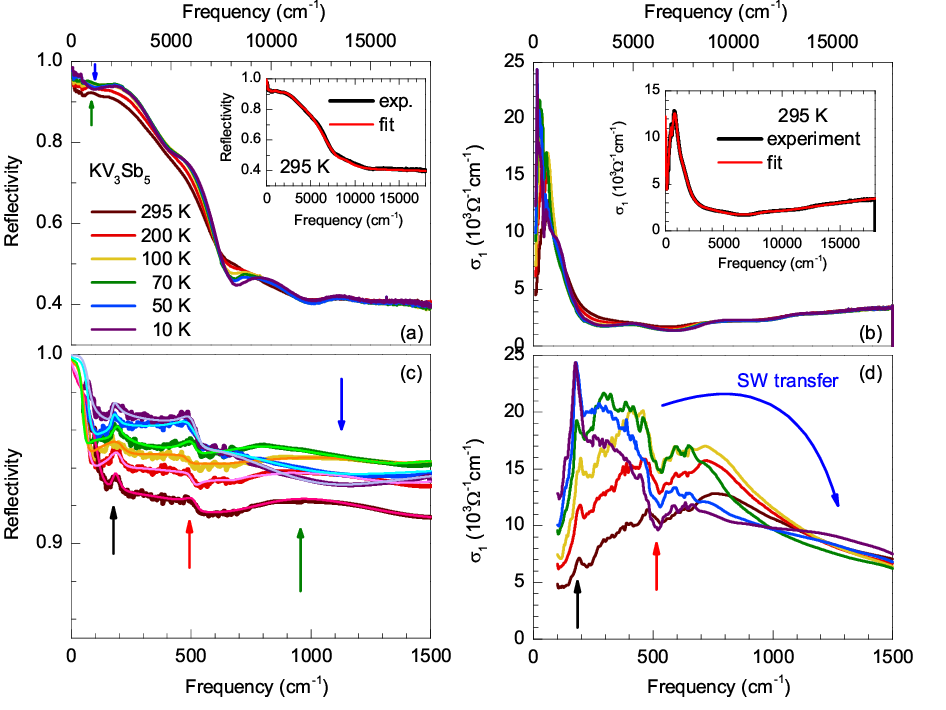}
	\caption{{\bf a, b} Frequency-dependent reflectivity and optical conductivity for various temperatures. Insets show the experimental spectra (black curves) and the Drude+Lorentz fits to the spectra (red curves). Blue and green arrows in {\bf a} represent the depletion due to density wave formation and dominant interband transitions at low energies, respectively. {\bf c} Low-energy reflectivity with Drude+Lorentz fits. Black and red arrows highlight the phonon and antiresonance features. Sharp upturn of the spectra at energies below 200~\cm\ signifies the existence of a very sharp Drude-component.  {\bf d} Corresponding low-energy optical conductivity, where a shifted spectra have been given in the main text Fig.2a.}	
	\label{RefOC}
\end{figure}

In Supplementary Figs.~\ref{RefOC}a,b temperature-dependent reflectivity and corresponding optical conductivity is given in a broad energy range. Simultaneous fit to the reflectivity and optical conductivity has also been given for room temperature, demonstrating the sharp Drude-component at low energies. Generally, high reflectivity corporates with the highly metallic nature of the compound. A broad absorption feature is also visible in bare reflectivity spectrum demonstrating the dominant effects of the interband transitions at these low energies (marked with green arrow). With decreasing the temperature below \Tc, density wave gap opens also seen as a depletion in reflectivity marked with the blue arrow in Supplementary Fig.~\ref{RefOC}a.

In Supplementary Figure~\ref{RefOC}c and d, we focus to the low energy regime, where the phonon mode and the high frequency antiresonance can be recognized more clearly. These features have been marked with black and red arrows, correspondingly. The low-energy optical conductivity is also given in the main text Fig.2a with shifted spectra.

\section*{Supplementary Note 3. Fit of the spectra}

Different contributions to the optical conductivity were modeled with the phenomenological Drude-Lorentz approach. They are presented in Fig.~2b in the main text along with the experimental spectrum. For the high-energy 482\,\cm\ mode we employed the asymmetric (Fano) line shape as described in the next sections.
\begin{align}
\label{Eps}
&\tilde{\varepsilon}=\varepsilon_1 + i\varepsilon_2,\\\notag
&\tilde{\varepsilon}(\omega)= \varepsilon_\infty - \frac{\omega^2_{p,{\rm Drude}}}{\omega^2 + i\omega/\tau_{\rm\, Drude}} + \sum\limits_j\frac{\Omega_j^2}{\omega_{0,j}^2 - \omega^2-i\omega\gamma_j}.
\end{align}
Here, $\varepsilon_\infty$ stands for high-energy contributions to the real part of the dielectric permittivity, whereas $\omega_{p,{\rm Drude}}$ and $1/\tau_{\rm\,Drude}$ are the plasma frequency and the scattering rate of the itinerant carriers, respectively. $\omega_{0,j}$, $\Omega_j$, and $\gamma_j$ describe the resonance frequency, width, and the strength of the $j^{th}$ excitation. The complex optical conductivity is then obtained as
\begin{align}
&\tilde{\sigma}=\sigma_1 + i\sigma_2,\notag\\
&\tilde{\sigma}(\omega)= -i\omega[\tilde{\varepsilon}-\varepsilon_\infty]/4\pi.
\label{Cond}
\end{align}

The phonon anomalies discussed in the main text suggest a strong electron-phonon coupling in \KVS. The lower 188\,\cm\ mode can be represented with a single Lorentzian, while the higher 482\,\cm\ mode is better reproduced with a Fano-like response:

\begin{equation}
\sigma_1(\omega) = \frac{\Delta\epsilon\omega^2\omega_0^2\gamma}{4\pi[(\omega^2-\omega_0^2)^2+\gamma^2\omega^2]}
 \label{lorentz}
\end{equation}

\begin{equation}
\sigma_1(\omega) = \frac{\Delta\epsilon\omega\gamma[\gamma\omega(q^2-1)^+2q(\omega^2-\omega_0^2)]}{4\pi[(\omega^2-\omega_0^2)^2+\gamma^2\omega^2]}
 \label{fano}
\end{equation}

Here, $\omega_0$, $\Delta\epsilon$, and $\gamma$ correspond to the resonance frequency, intensity, and the linewidth of the phonon mode, respectively. $q$ is the dimensionless coupling constant that describes the asymmetry of the mode. Here, it also takes negative values as the high-energy mode is a strong antiresonance. 

\section*{Supplementary Note 4. Spectral weight analysis}

Localization peak has been extracted from the Drude-Lorentz fits of the optical conductivity throughout the measured temperatures. An example of such a fit is given in the main text (Fig.~2b). In Fig.~2f, we demonstrate the temperature evolution of the unconventional carriers. The peak positions have been marked with the solid circles that illustrate the red shift upon cooling. The spectral weight (SW) of the localization peak is given in Fig.~2g and has been calculated with the general relation:

\begin{equation}
 {\rm SW} = \int_0^{\omega_{c}}\sigma_1(\omega)d\omega
\end{equation}

Here, $\omega_c$ is the cut-off frequency $\sim$~2\,eV, and $\sigma_1(\omega)$ is taken from Fig.~2f obtained through the fit of the whole spectra. The reduction upon \Tc\ is visible and indicates the SW redistribution between the localization peak and the high-energy absorption. 

To confirm the proposed distribution, the overall spectral weight (Fig.~2g) has also been calculated in a similar manner from the measured optical conductivity curves. As expected from the optical sum rules, the overall SW is conserved within the measurement range of $\sim$~2\,eV.

\section*{Supplementary Note 5. Plasma frequency }

Here we show the experimental estimate of the plasma frequency via the real part of the dielectric permittivity, $\epsilon_1$, for \KVS. As given in Supplementary Fig.~\ref{plasma}a, the negative values and the divergence of the permittivity towards $\omega\rightarrow 0$ demonstrate the highly metallic nature of \KVS. The plasma frequency is screened by the high energy interband transitions, where the zero-crossing of $\epsilon_1$ reflects the screened plasma frequency, $\omega_{\rm p}^{\rm screened}$. However, the unscreened plasma frequency can be estimated as $\omega_{\rm p} = \omega_{\rm p}^{\rm screened} \sqrt{\epsilon_\infty}$, where $\epsilon_\infty$ is determined by the high-energy contributions and stays around 5 at all temperatures as shown in Supplementary Fig.~\ref{plasma}a. The zero-crossing in normal state is at $\sim$~0.95~eV (7680~\cm), allowing us to estimate the plasma frequency of around 2.12~eV. It does not change significantly across the density wave transition as depicted in Supplementary Fig.~\ref{plasma}b.

\begin{figure}
	\centering
	\includegraphics[width=0.8\columnwidth]{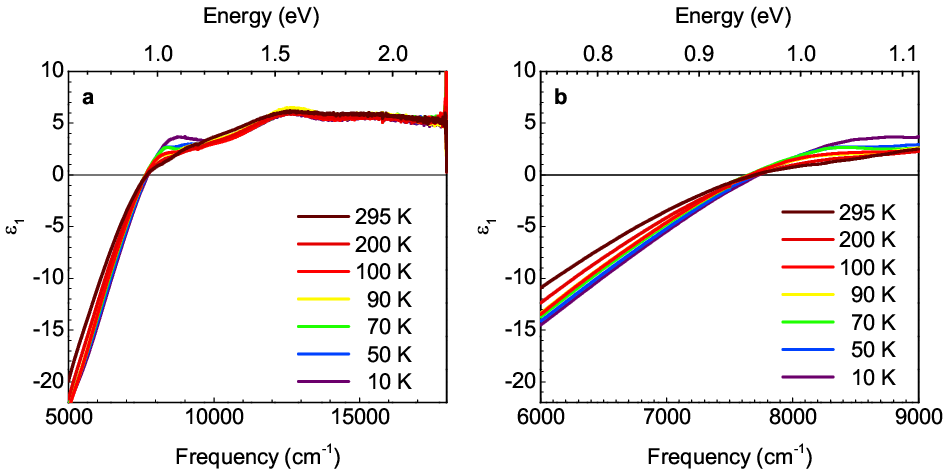}
	\caption{{\bf a} Frequency-dependent real part of the dielectric permittivity for selected temperatures above and below the density wave transition. $\epsilon_\infty$ is estimated to be around 5 for all the temperatures. {\bf b} Zero-crossing of $\epsilon_1$ (screened plasma frequency) shown to be around 0.95~eV.}		
	\label{plasma}
\end{figure}

Alternatively, one can estimate the plasma frequency through the Drude-Lorentz fits. Here the Drude-contribution and the localization peak give rise to an effective plasma frequency, $\omega^2_{\rm p,effective} = \omega^2_{\rm p,Drude2} + \omega^2_{\rm p,localization}$ , and amounts to 1.99 eV. A small mismatch with the zero-crossing of $\epsilon_1$ is due to the difficulty of properly estimating the Drude contribution, as only the tail is visible. With this assumption, the correlation ratio is around 0.33, which agrees reasonably with the value of 0.37 given in the main text.

\section*{Supplementary Note 6. Additional computational results}

\subsection{Fermi energy and $k$-mesh convergence. }

\begin{figure}[h!]
	\centering
	\includegraphics[width=0.8\columnwidth]{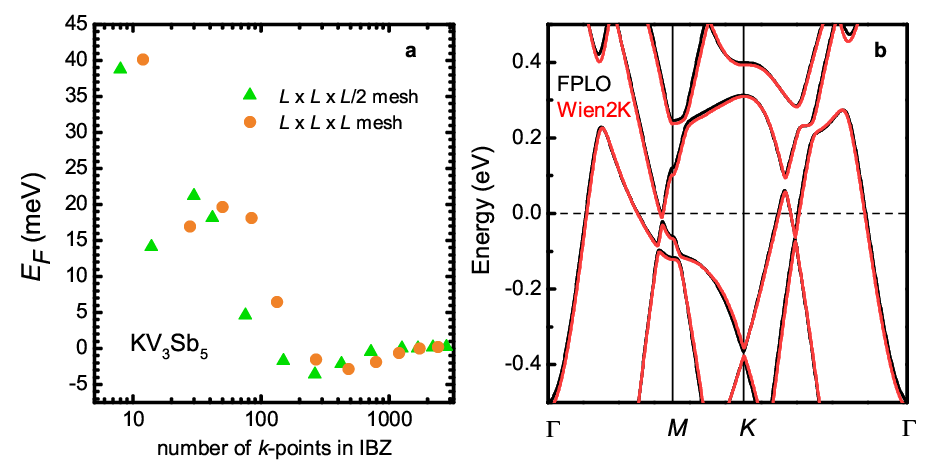}
	\caption{{\bf a} Position of the Fermi energy ($E_F$) for the undistorted KV$_3$Sb$_5$ structure vs. the number of $k$-points in the irreducible part of the first Brillouin zone (IBZ). The converged value of $E_F$ is taken as zero. Our calculations of the band dispersion and optical conductivity are performed on the $L\times L\times L/2$ mesh with at least 2000 points ($L\geq 36$) where good convergence is reached. {\bf b} Comparison of the band structures calculated in \texttt{FPLO} and \texttt{Wien2K}. }	
	\label{k-mesh}
\end{figure}

We observed that the position of the Fermi level ($E_F$) is somewhat dependent on the choice of the $k$-mesh, as shown in Supplementary Fig.~\ref{k-mesh}a. The smaller number of $k$-points leads to an overestimate of $E_F$ by $20-40$~meV, especially if the $L\times L\times L$ mesh not matching the anisotropy of the crystal structure is used.  To ensure the accuracy and self-consistency of our calculations, we used the $L\times L\times L/2$ mesh and the ample number of $k$-points that allowed the convergence of $E_F$ of better than 2~meV. We have also cross-checked that different DFT codes, \texttt{FPLO} and \texttt{Wien2K}, lead to nearly indistinguishable band structures with the same position of the Fermi level, see Supplementary Fig.~\ref{k-mesh}b.

\subsection{Optical conductivity.}

Supplementary Figure~\ref{DFT} shows optical conductivity of the undistorted \KVS{} structure calculated for different doping levels within DFT (panel a) and for the stoichiometric \KVS{} within DFT+$U$ (panel b). Low-energy part of the spectrum is dominated by two peaks around 0.1\,eV and 0.25\,eV due to the B--C and \mbox{B--D} transitions, respectively. Relative intensities of these peaks strongly depend on the position of the Fermi level and on adding correlations within DFT+$U$. Best agreement with the experimental optical conductivity is obtained in DFT (without $U$) upon shifting the Fermi level by $+64$\,meV. This chemical potential nominally corresponds to the doping level of $0.4e$/f.u. A similar suppression of the 0.25\,eV peak and enhancement of the 0.1\,eV peak are observed in DFT+$U$ with the on-site Coulomb repulsion $U=2$\,eV, Hund's exchange $J=0$\,eV, and double-counting correction in the atomic limit. In this case, a weak magnetic moment of 0.07 $\mu_B$ on vanadium atoms is obtained. Both methods seem to yield approximations to the real \KVS{} band structure, where band energies around the $M$-point are renormalized compared to the DFT prediction. We note that ARPES results of Ref.~\onlinecite{Yang2020}, the only spectroscopic study of \KVS{} available so far, also show a slight mismatch between the calculated and experimentally measured bands along the $\Gamma-K$ direction. Better agreement can be reached by the upward shift of the Fermi level by about 50\,meV, which is compatible with the shift required to reproduce the experimental optical conductivity obtained in our study.

For the DW state, atomic displacements should be taken into account, because the parent \KVS{} structure is unstable in DFT, whereas scanning tunneling microscopy experiments\cite{Jiang2021} also suggest the superstructure formation. Therefore, we performed calculations for the star and hexagon DW structures. Their optical conductivity did not show any significant dependence on the doping level and $U$ value. In this case, the absorption peak around 0.2\,eV is observed systematically, in agreement with the experiment (Fig.~3e). Therefore, the 0.15-0.2\,eV peak in the DW state is related to the structural distortion, thus supporting the formation of the density-wave state in \KVS{}.

\begin{figure}
	\centering
	\includegraphics[width=0.5\columnwidth]{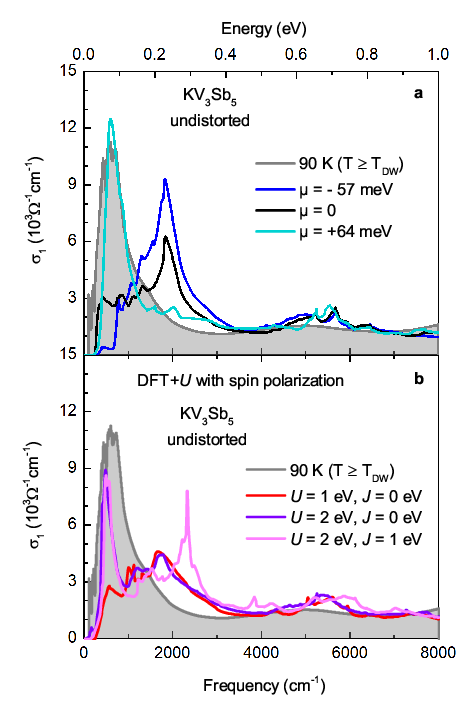}
	\caption{{\bf Comparison of the experimental interband contribution to the optical conductivity at 90\,K (above the DW transition) with the DFT results for the undistorted crystal structure.} {\bf a} for different hole and electron dopings expressed by the chemical potential $\mu$, and {\bf b} for DFT+$U$ calculations at $\mu=0$. Electron doping leads to a suppression of the peak around 0.25\,eV and enhances the peak around 0.1\,eV, resulting in the best match to the experiment. Similar effect can be reproduced with including $U$ on the DFT+$U$ level. }		
	\label{DFT}
\end{figure}

\subsection{Possible spin states of DW structures.}

\begin{figure}
	\centering
	\includegraphics[width=1\columnwidth]{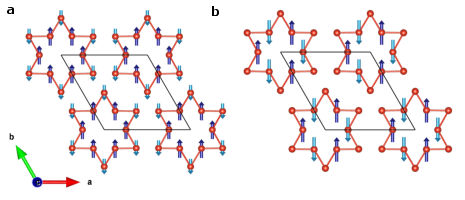}
	\caption{{\bf SDW states of the star structure.} {\bf a} type-A SDW with parallel spins on the inner hexagon of the star; {\bf b} \mbox{type-B} SDW with antiparallel spins on the inner hexagon. The \mbox{type-B} SDW breaks 6-fold symmetry of the charge bond order, increases local moment, and lowers total energy of the system. The high stability of this state indicates the proclivity of the star structure to the formation of a conventional spin-density wave. }		
	\label{star}
\end{figure}

Calculations on the DFT level (without adding $U$) suggest that both hexagon and star structures (Fig.~3) should be non-magnetic. Local magnetic moments can be nevertheless introduced in DFT+$U$ in the form of a conventional spin-density wave (SDW) where spin density is modulated throughout the crystal. Although different from spin bond order of the kagome Hubbard model, this is the best approximation that density-functional methods can offer. Here, we discuss such SDW's that may appear in the charge-ordered hexagon and star structures. The values of energy and magnetic moment are quoted for $U=2$\,eV and $J=0$\,eV in DFT+$U$.

Two types of SDW, A and B, are possible in the star structure. Type-A SDW preserves six-fold symmetry and entails parallel spins on the inner hexagon of the star along with opposite spins on the outer sites of the star (Supplementary Fig.~\ref{star}a). Slightly different magnetic moments of 0.85\,$\mu_B$ and $-1.0$\,$\mu_B$, respectively, result in a net ferrimagnetic state. It is less stable than the antiferromagnetic state, type-B SDW with antiparallel spins on the inner hexagon (Supplementary Fig.~\ref{star}b). It features higher magnetic moments of $1.3-1.4$\,$\mu_B$ and lies 50\,meV/f.u. lower in energy compared to type-A SDW, which is in turn 35\,meV/f.u. lower in energy than the non-magnetic state of the star structure. The total stabilization energy of the SDW is thus 85\,meV/f.u. for the given choice of the DFT+$U$ parameters. Therefore, electronic correlations are likely to introduce local magnetism into the star structure and give rise to an antiferromagnetic type-B SDW that breaks six-fold symmetry of charge bond order.

Hexagon structure is very different. It does not support type-B SDW at all, whereas type-A SDW involves only tiny magnetic moments of 0.05\,$\mu_B$ and shows no significant energy gain compared to the non-magnetic state (the energy gain due to the SDW is 0.5\,meV/f.u. only). This result indicates that hexagon structure is not prone to the formation of conventional SDW's, which then allows embedding spin degrees of freedom in a different form, such as spatial modulation of spin currents in the spin bond order (triplet $p$-wave) state of the kagome Hubbard model.

We have also checked starting configurations with non-collinear spins. In this case, DFT+$U$ calculations converged to one of the aforementioned solutions or to the non-magnetic solution.

\section*{Supplementary References}

\end{document}